# Three-dimensional intraoperative light field fluorescence imaging system for PpIX-guided tumour resection

Yucheng Bian[a], Sebastien Ourselin[a], Yijing Xie[a]*

a. Department of Surgical and Interventional Engineering, School of Biomedical Engineering & Imaging Sciences, King's College London, London, United Kingdom

## ABSTRACT

Conventional 2D fluorescence imaging in glioma surgery cannot separate intrinsic fluorophore strength from attenuation with depth, creating depth-intensity ambiguity that can compromise assessment of residual tumour and fluorescence-based grading. This study develops and validates a dual mode light field imaging system that could capture 3D structure and depth corrected fluorescence in a single snapshot by adapting a commercial Lytro Illum camera. A custom 3D printed depth standard was used to optimise main lens focal length and to derive a grayscale-distance linearity from Lytro Desktop depth maps. CdSe/ZnS quantum dot targets and fluorescent brain phantoms were imaged to establish fluorescence intensity distance attenuation models and to recover intrinsic fluorescence. In system optimization, the increasing FU strengthened grayscale depth linearity and achieved millimetre scale vertical resolution ($R^2 > 0.95$) for FU $\geq$ 60 mm. Higher concentration quantum dot wells (levels 3 to 5) of the fluorescent target showed consistent attenuation. In fluorescence mode, the deviations of distance estimations across six regions of a fluorescent brain phantom were 0.14 to 2.45% with intensity prediction errors from −11.73% to 6.08% based on the fluorescence intensity-distance model, enabling recovery of intrinsic quantum dot concentrations which are mimicking PpIX characteristics in glioma. This research supports light field imaging as a practical approach for depth resolved quantitative fluorescence and improved intraoperative tumour characterisation.

## 1. INTRODUCTION

Primary brain tumours pose major challenges due to their infiltrative growth patterns and proximity to eloquent brain regions [1-2]. Surgical resection remains the cornerstone treatment for malignant tumours such as gliomas, and the extent of resection strongly correlates with overall and progression-free survival [3-4]. However, tumour margins are often indistinct under conventional white-light microscopy, and small errors in boundary identification can lead to residual tumour or postoperative neurological deficits, driving the need for reliable intraoperative visualization [5-9]. Fluorescence-guided surgery (FGS), enabled by agents such as 5-aminolevulinic acid (5-ALA) or indocyanine green (ICG), improves tumour conspicuity and margin detection and has become an important component of modern neurosurgical workflows [10-12].

Despite these advances, most clinical fluorescence systems remain wide-field and two-dimensional. Quantitative and spectrally resolved approaches, including frequency-domain fluorescence imaging and multispectral fluorescence imaging, can improve discrimination but remain planar measurements that are susceptible to depth-dependent attenuation [13-16]. Consequently, measured fluorescence intensity is confounded by source–camera distance, producing a clinically relevant depth–intensity ambiguity that can mislead intraoperative interpretation.

*Corresponding author: yijing.xie@kcl.ac.uk

To provide 3D context, optical surface-imaging techniques, particularly stereo vision, have been adopted for surgical navigation. Commercial stereo-vision exoscopes (e.g., ORBEYE, VITOM® EAGLE, and KINEVO) improve depth perception, but depth estimation can degrade in texture-poor scenes and under occlusion, and a dual-camera configuration increases system footprint and provides only a single baseline [17-23]. Other 3D modalities, including laser triangulation and structured light, can deliver high geometric accuracy but are constrained intraoperatively by specular reflections, occlusions, motion sensitivity, and scanning process [24-26].

Light-field imaging (LFI) captures spatial and angular information in a single exposure using a microlens array, enabling multi-view imaging, digital refocusing, and computational depth estimation [27-29]. Dense angular sampling offers the potential to mitigate occlusions and supports snapshot 3D reconstruction without scanning, which is advantageous in confined neurosurgical workspaces [30-31]. Recent work has also begun to explore LFI for neurosurgical fluorescence quantification, motivating development toward depth-corrected fluorescence imaging [32].

In this study, a dual-mode neurosurgical LFI system was realised by adapting a Lytro Illum plenoptic camera with spectral filtering and dedicated white-light and blue-excitation illumination. System performance was characterised using a customed depth standard and fluorescent brain phantoms, and the main-lens focal length and working distance were optimised to achieve a neurosurgically relevant field of view, depth of field, and vertical resolution. A linear relation between Lytro depth-map grayscale values and physical distance enables 3D reconstruction without proprietary microlens parameters, while modelling distance-dependent fluorescence attenuation supports depth-corrected recovery of intrinsic fluorescence intensity, thereby reducing the depth–intensity ambiguity of conventional 2D fluorescence imaging.

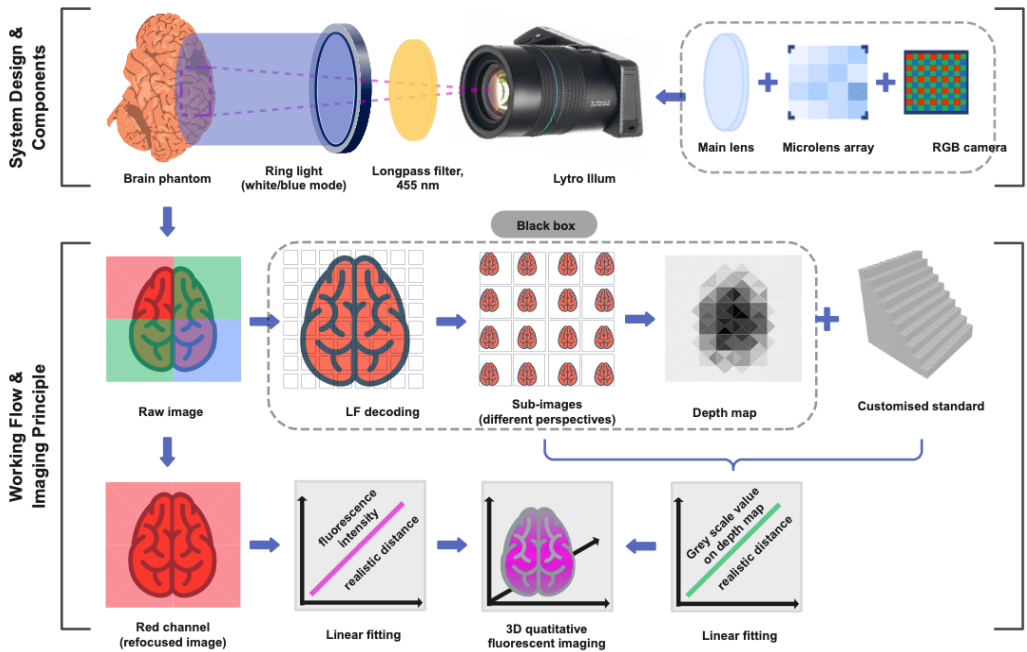

Fig. 1 System architecture, light-field imaging principle, and fluorescence quantification workflow of the neurosurgical light-field imaging system based on the Lytro Illum platform.

## 2. METHODS

### 2.1 Design of the fluorescence light field imaging system

The customised dual-mode light-field fluorescence imaging system incorporated a commercial plenoptic camera (Lytro Illum), an longpass emission filter with a cutoff wavelength at 455 nm, and a coaxial ring illuminator providing white-light or blue-excitation illumination (Fig. 1b). The Lytro Illum adopts a plenoptic 2.0 design with an adjustable main lens and a fixed microlens array (MLA) above the sensor, sampling spatial and angular information for disparity-based depth estimation from sub-aperture views. In this work, imaging parameters were systematically optimised for intraoperative neurosurgical requirements, targeting a clinically relevant field of view (FOV), sufficient depth of field (DoF), and millimetre vertical resolution. Because the MLA geometry is fixed, DoF and axial resolution are governed primarily by main-lens focal length and working distance: larger focal lengths and shorter distances typically narrow DoF but improve axial discrimination. These two parameters were therefore explored experimentally to maximise depth fidelity while preserving adequate anatomical coverage, and the final configuration was validated using the customised stage standard.

## 2.2 Customised stage standard development

A customised 3D-printed PLA staircase target was fabricated to quantify system FOV, DoF, and vertical resolution (Fig. 1b). Each row contains ten 1-mm steps (continuous across rows), and each column contains ten 10-mm steps, providing a 100-mm axial span for DoF assessment. Each step includes a 6-mm-diameter, 1-mm-deep well filled with pigment/QD mixtures and levelled with silicone or resin to form a planar surface flush with the step height.

## 2.3 Quantified 3D reconstruction principle and system performance assessment

Because Lytro Illum's intrinsic parameters (focal length of microlength and main lens to MLA spacing) are proprietary, conventional sub-aperture disparity models cannot recover absolute depth. Instead, the depth map exported on Lytro Desktop was used, leveraging its reported grayscale–distance linearity within 50–600 mm working distance [32]. Grayscale values were sampled and fitted to ground-truth distances to obtain a linear-fitting model that converts depth grayscale to physical distance, and its repeatability and reliability was assessed across white-light and fluorescence modes. The customised stage target was then imaged to test millimetre-scale axial separability by ROI analysis on 1mm steps. Finally, focal length and working distance were tuned to meet neurosurgical requirements of FOV, DoF, and vertical resolution.

## 2.4 Quantified 3D reconstruction principle and system performance assessment

A commercial DoF calibration target with a certified 50 mm depth span was imaged under white light to assess the accuracy of depth reconstruction. Grayscale values at the nearest and farthest target surfaces were extracted from the depth map and converted to axial positions using the established grayscale–distance linearity, yielding an estimated depth. Reconstruction deviation was quantified as

$$Deviation = \frac{|d_{est} - d_{true}|}{d_{true}}$$

$d_{est}$, represents the estimated distance by the LF system, and $d_{true}$ represents the realistic distance.

Fluorescence-mode reconstruction was evaluated on a silicone brain phantom with fluorescent regions. The fluorescence was provided by CdSe/ZnS quantum dots, selected as stable surrogates with PpIX-like characteristics [34]. The grayscale–distance line relationship under fluorescence mode was applied to estimate the distances between system and fluorescent regions and the deviation was also computed to quantify reconstruction reliability under fluorescence illumination.

## 2.5 3D quantified fluorescence mapping

A 5×5 fluorescence calibration phantom was constructed to jointly quantify fluorescence intensity and 3D position. Each row contained CdSe/ZnS QD wells at 0.0001–0.0040 mg/mL and was imaged under blue-excitation. To derive fluorescence intensity–distance attenuation and define the usable fluorescence working range, the phantom was positioned from 180 to 530 mm in 25 mm steps. For each distance, mean fluorescence was extracted from ROIs within the frontmost row of wells in the red channel of refocused RGB images, generating intensity–distance curves for each concentration. Model generalizability was assessed on a silicone brain phantom with localized QD inclusions matched to the standard. Finally, combining the LF grayscale-to-depth calibration with the fluorescence attenuation model enabled estimation of absolute 3D position and distance-compensated intrinsic fluorescence (QD concentration). Such correction can mitigate depth-related bias and may support intraoperative fluorescence-based glioma grading.

## 3. RESULTS

**3.1 System parameter opimisation and the grayscale - depth linearity**

Under As shown in Fig. 2(a), the customised depth standard was fabricated, and its wells were filled with RGB pigment resin while remaining flush with each step. To study the effect of main-lens focal length on DoF and vertical resolution, the target was placed 50 mm from the camera and imaged at $FU$= 30 – 70 mm. Depth-map grayscale values were extracted from well ROIs and fitted against 1mm step indices (Fig. 2(c)). The linearity improved with increasing $FU$ reaching $R^2$> 0.95 at $FU$=60 and 70 mm (Fig. 2(d)), indicating millimetre-level axial discrimination. Under white-light imaging, the RGB depth target showed strong linearity (Fig. 2(e), $R^2$= 0.981) over 13 – 58 mm, supporting an effective DoF ≥ 45 mm and non-overlapping ROI distributions across 1-mm increments. Under fluorescence mode, a QD-based depth target (Fig. 2(b)) also exhibited strong grayscale–distance linearity (Fig. 2(f), $R^2$ = 0.993), with an effective DoF of at least 35 m.

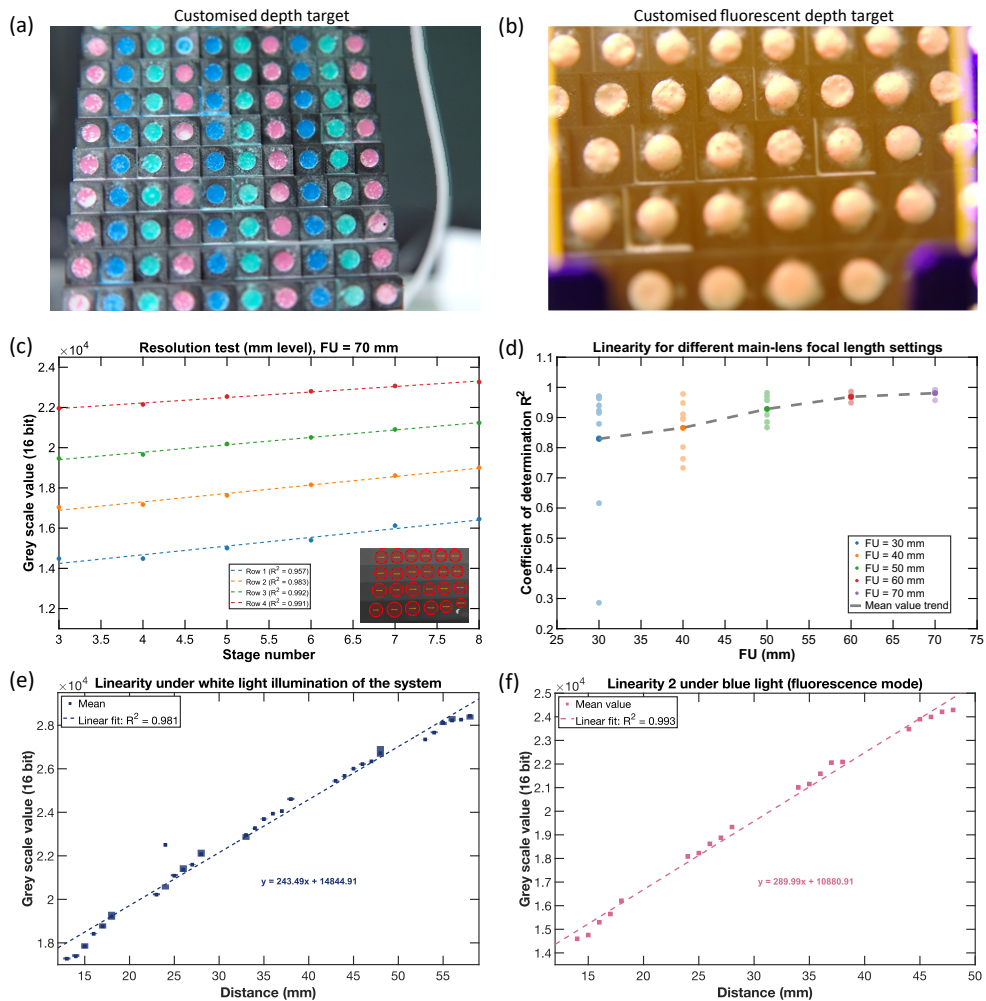

Fig.2 (a) Customised depth standard filled with resin mixtures containing red, green, and blue pigments. (b) Customised fluorescent depth standard filled with resin mixtures containing CdSe/ZnS quantum dots of uniform concentration. (c) Linear relationships between the grayscale values extracted from the depth-map ROIs of wells in the same rows and the corresponding stage numbers (each stage differing by 1 mm in height) when $FU$=70 mm. (d) Variation of the coefficient of determination $R^2$ as a function of $FU$, illustrating the improvement in linearity with increasing focal length. (e) Grayscale-distance linearity under white light illumination of the system. (f) Grayscale-distance linearity in fluorescence mode.

### 3.2 Validation of depth estimation

Based on the grayscale-distance linearity on fluorescence mode, the distance reconstruction was further assessed on the brain phantom containing embedded fluorescent regions under blue excitation as show in Fig. 3 (b). The fluorescent areas contain the QD concentrations 3–5 on fluorescence stage target, as shown in Fig. 3(a). The estimated distances of the six fluorescent regions (A–F) were calculated as shown in Table 1. The depth-estimation deviation across these regions ranged from 0.14% to 2.45%, demonstrating that the LF-based reconstruction approach maintains reliable accuracy on anatomically realistic phantoms.

Table 1. Light-field–based distance estimations for six fluorescent regions on the brain phantom.

|  | Region A (conc. 4) | Region B (conc. 3) | Region C (conc. 5) | Region D (conc. 5) | Region E (conc. 3) | Region F (conc. 4) |
|---|---|---|---|---|---|---|
| Realistic distance (mm) | 330.40 | 351.70 | 322.30 | 322.94 | 310.61 | 390.49 |
| Estimated distance (mm) | 338.50 | 352.20 | 329.61 | 325.25 | 314.67 | 314.43 |
| Spatially deviation % | 2.45 | 0.14 | 2.27 | 0.72 | 1.31 | 1.59 |

### 3.3 Depth-resolved fluorescence quantification

To quantify distance-dependent fluorescence attenuation and enable depth-resolved fluorescence mapping, fluorescence stage targets with CdSe/ZnS QD wells (0.0001–0.0040 mg/mL) were fabricated as shown in Fig. 4(a). The front row of the stage target was imaged from 180 mm with 25-mm increments , and the mean fluorescence was extracted from red-channel ROIs in refocused images as shown in Fig. 4(c). Concentration-3 wells showed fluorescence intensity–distance linearity over 180–430 mm, while concentrations 4–5 were linear over 255–530 mm, with similar attenuation slopes (–0.450, –0.453, –0.462), indicating consistent decay in fluorescence signal and distance.

The fluorescent brain phantom was then imaged within 255–430 mm. Using the estimated distances in the Table 1and the intensity–distance models, fluorescence intensities were predicted as red points in Fig. 3(e), red) and compared with extracted ROI values in red channel (black points in Fig. 3(d), yielding deviations of −11.73% to 6.08%. Finally, with the estimated distances from Light field camera, the distance-based correction recovered intrinsic fluorescence levels for area A–F based on the fluorescence-distance attenuation, enabling concentration inference from a single LF acquisition, as shown in Fig. 3(f).

Based on the estimated distances in Table 1 and the fluorescence intensity–distance linear models derived from concentration-3 to concentration-5 wells in Fig. 3(c), the predicted fluorescence intensities for regions A–F were calculated as shown as red markers in Fig. 3(e). Fluorescence intensities were extracted from the red-channel ROIs as ground truth corresponding to regions A–F in the refocused image, as shown in Fig. 3(d). The predicted intensities (red markers) closely matched the experimentally measured values (black markers), with intensity-prediction deviations ranging from −11.73% to 6.08%, as shown in Fig. 3(e).

Using the LF-estimated distances together with the fluorescence intensity–distance linear fitting, the measured fluorescence signals were corrected for geometric attenuation, thereby recovering the intrinsic fluorescence strength of each region. The restored fluorescence levels for regions A–F are shown in Fig. 3(f), illustrating that the method successfully compensates for distance-dependent signal decay and allows fluorophore concentration to be quantitatively inferred from a single LF acquisition.

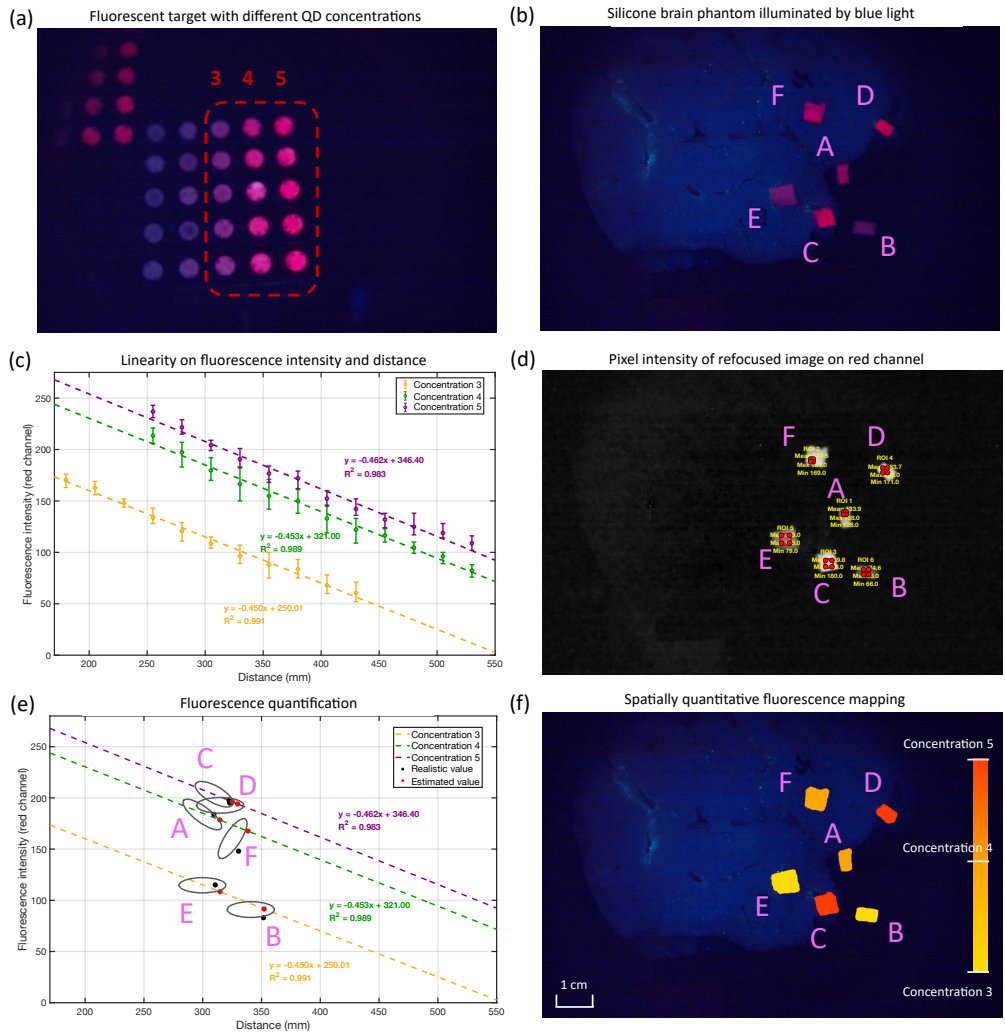

Fig.3 (a) A fluorescent depth target embedded with resin mixtures containing five different QD concentrations (increasing from left to right along the first row). Under blue-light excitation, the wells emit fluorescence intensities proportional to their concentrations. (b) Fluorescence-mode image of the same brain phantom acquired under blue-light excitation. (c) Linear fits between fluorescence intensity and actual distance for wells with QD concentrations 3–5. Saturated regions and nonlinear segments outside the camera's dynamic range were excluded from the regression. (d) Fluorescence intensities extracted from the red channel of the refocused image for regions A–F. (e) Estimated fluorescence intensities (red markers), calculated using the grayscale-derived distances and the fluorescence intensity–distance linear models, plotted alongside the measured fluorescence intensities extracted from the red channel (black markers). Predicted and measured values for each region are grouped using dashed circles. (f) Intrinsic QD concentrations (fluorescence levels) recovered for each region by combining the LF-estimated distances with the fluorescence intensity–distance linearity.

# 4. DISCUSSION

## 4.1 System parameter optimization

A neurosurgical light-field imaging system based on Lytro Illum was developed and optically optimised for intraoperative 3D surface reconstruction. In the testings, it showed that increasing *FU* strengthened the grayscale–depth linearity and enabled millimetre-scale vertical resolution, with *FU* =60 or 70 mm achieving 1mm depth discrimination. This trend is consistent with plenoptic sampling. A fixed MLA angular sampling means that a narrower DoF at larger focal length increases axial sampling density [27,35].

A linear mapping between Lytro Desktop depth-map grayscale values and physical distance was established to enable metrically accurate 3D reconstruction. Calibrated with a customised depth target, the grayscale–distance relationship remained reliable in both white-light and fluorescence modes, showing stability to illumination changes and strong reproducibility under blue excitation.

## 4.2 Quantitative fluorescent mapping

A fluorescence target with five CdSe/ZnS QD concentrations was fabricated to model fluorescence attenuation versus distance. Imaging across distances showed that higher concentrations (3–5) followed consistent intensity–distance linearity with similar slopes within their working ranges. The spatial quantification workflow was validated on a silicone brain phantom containing regions with concentrations 3–5. Thus, this accurate QD localisation suggests potential for intraoperative localisation of PpIX-expressing tumour regions to support fluorescence-guided glioma surgery. The phantom and fluorescent stage target contain a PpIX-mimicking fluorophore, CdSe/ZnS quantum dots, reported to enable to provide PpIX fluorescence characteristics in glioma [34]. Thus, this accurate QD localisation suggests potential for intraoperative localisation of PpIX-expressing tumour regions to support fluorescence-guided glioma surgery.

With grayscale-distance linearity, it provided distance estimations with 0.14–2.45% deviation, supporting reliable localisation. Using these estimated distances and the fitted attenuation models, fluorescence intensities were predicted and compared with red-channel values extracted from refocused images, yielding deviations of −11.73% to 6.08%. Combining LF-derived distances with the attenuation model for the correction of geometric signal loss and recovery of intrinsic QD concentration, the system enables depth-corrected fluorescence quantification for intraoperative neurosurgical surface mapping.

## 4.1 Limitations and future work

Because key plenoptic parameters (microlens focal length, MLA–sensor spacing, main-lens geometry) are unavailable, physically disparity-to-depth reconstruction is not feasible. Therefore, the depth estimation relies on an empirical grayscale–distance model from Lytro Desktop, which may not generalise beyond this Lytro-based design. In fluorescence quantification, low-QD wells (concentrations 1–2) did not have reliable fluorescence intensity-distance linearity due to limited sensor sensitivity and dynamic range, indicating the need for a scientific-grade detector with higher quantum efficiency and wider dynamic range. Overall, these limitations motivate a neurosurgery-tailored, open-architecture LF system co-designed via ray-tracing and advanced toward learning-enabled reconstruction [38,39]. In addition, future work should develop a phantoms with denser, clinically relevant concentration steps aligned with PpIX levels across glioma grades to validate intensity–distance correction.

# 5. CONCLUSIONS

This work demonstrates a fluorescence light-field imaging system for neurosurgery that enables simultaneous 3D reconstruction and depth-resolved fluorescence quantification. Using a customised depth standard, the system achieved millimetre-scale vertical resolution, extended depth of field, and a clinically relevant field of view. A grayscale–distance calibration, consistent in both white-light and fluorescence modes, supported accurate surface geometry recovery.

Fluorescence intensity–distance attenuation models derived from a QD stage standard enabled correction of distance-dependent signal loss and reconstruction of intrinsic fluorescence strength. Validation on anatomically realistic brain phantoms confirmed high spatial accuracy and quantitative agreement, addressing the depth–intensity ambiguity inherent to conventional 2D fluorescence imaging. Overall, the results support light-field imaging as a compact, computational approach for depth-corrected fluorescence assessment in glioma surgery, and provide a foundation for clinically optimised systems. Further improvements in optical design, sensor sensitivity, and fluorescence detection limits are needed for translation.

## ACKNOWLEDGMENTS

This work was supported by the Royal Academy of Engineering [RF/2122/21/354] and by core funding from the EPSRC DTP [EP/W524475/1].

## REFERENCES

[1] Patel, Vimal, and Vishal Chavda. "Intraoperative glioblastoma surgery-current challenges and clinical trials: An update." Cancer pathogenesis and therapy vol. 2,4 256-267. 2 Dec. 2023, doi:10.1016/j.cpt.2023.11.006

[2] Ilic, Irena, and Milena Ilic. "International patterns and trends in the brain cancer incidence and mortality: An observational study based on the global burden of disease." Heliyon vol. 9,7 e18222. 13 Jul. 2023, doi:10.1016/j.heliyon.2023.e18222

[3] McGirt, Matthew J et al. "Extent of surgical resection is independently associated with survival in patients with hemispheric infiltrating low-grade gliomas." Neurosurgery vol. 63,4 (2008): 700-7; author reply 707-8. doi:10.1227/01.NEU.0000325729.41085.73

[4] Price, Mackenzie et al. "CBTRUS Statistical Report: Primary Brain and Other Central Nervous System Tumors Diagnosed in the United States in 2017-2021." Neuro-oncology vol. 26,Supplement_6 (2024): vi1-vi85. doi:10.1093/neuonc/noae145

[5] Elhawary, Haytham et al. "Intraoperative real-time querying of white matter tracts during frameless stereotactic neuronavigation." Neurosurgery vol. 68,2 (2011): 506-16; discussion 516. doi:10.1227/NEU.0b013e3182036282

[6] Zhang, Zoe Z et al. "The Art of Intraoperative Glioma Identification." Frontiers in oncology vol. 5 175. 30 Jul. 2015, doi:10.3389/fonc.2015.00175

[7] Stummer, Walter et al. "Extent of resection and survival in glioblastoma multiforme: identification of and adjustment for bias." Neurosurgery vol. 62,3 (2008): 564-76; discussion 564-76. doi:10.1227/01.neu.0000317304.31579.17

[8] Rees, J H. "Diagnosis and treatment in neuro-oncology: an oncological perspective." The British journal of radiology vol. 84 Spec No 2,Spec Iss 2 (2011): S82-9. doi:10.1259/bjr/18061999

[9] Young, Richard M et al. "Current trends in the surgical management and treatment of adult glioblastoma." Annals of translational medicine vol. 3,9 (2015): 121. doi:10.3978/j.issn.2305-5839.2015.05.10

[10] Stummer, Walter et al. "Fluorescence-guided surgery with 5-aminolevulinic acid for resection of malignant glioma: a randomised controlled multicentre phase III trial." The Lancet. Oncology vol. 7,5 (2006): 392-401. doi:10.1016/S1470-2045(06)70665-9

[11] Schupper, Alexander J et al. "Fluorescence-Guided Surgery: A Review on Timing and Use in Brain Tumor Surgery." Frontiers in neurology vol. 12 682151. 16 Jun. 2021, doi:10.3389/fneur.2021.682151

[12] Rodriguez, Benjamin et al. "Fluorescence-Guided Surgery for Gliomas: Past, Present, and Future." Cancers vol. 17,11 1837. 30 May. 2025, doi:10.3390/cancers17111837

[13] Xie, Yijing et al. "Wide-field spectrally resolved quantitative fluorescence imaging system: toward neurosurgical guidance in glioma resection." Journal of biomedical optics vol. 22,11 (2017): 1-14. doi:10.1117/1.JBO.22.11.116006

[14] Black, D., Kaneko, S., Walke, A. et al. Characterization of autofluorescence and quantitative protoporphyrin IX biomarkers for optical spectroscopy-guided glioma surgery. Sci Rep 11, 20009 (2021). doi:10.1038/s41598-021-99228-6

[15] Sibai, Mira et al. "Quantitative subsurface spatial frequency-domain fluorescence imaging for enhanced glioma resection." Journal of biophotonics vol. 12,5 (2019): e201800271. doi:10.1002/jbio.201800271

[16] Wirth, Dennis et al. "Feasibility of using spatial frequency-domain imaging intraoperatively during tumor resection." Journal of biomedical optics vol. 24,7 (2018): 1-6. doi:10.1117/1.JBO.24.7.071608

[17] Gosta, Miran and Mislav Grgic. "Accomplishments and challenges of computer stereo vision." Proceedings ELMAR-2010 (2010): 57-64.

[18] D. Scharstein, R. Szeliski and R. Zabih, "A taxonomy and evaluation of dense two-frame stereo correspondence algorithms," Proceedings IEEE Workshop on Stereo and Multi-Baseline Vision (SMBV 2001), Kauai, HI, USA, 2001, pp. 131-140, doi: 10.1109/SMBV.2001.988771.

[19] Zhou, Linglong et al. "A Comprehensive Review of Vision-Based 3D Reconstruction Methods." Sensors (Basel, Switzerland) vol. 24,7 2314. 5 Apr. 2024, doi:10.3390/s24072314

[20] Fiani, Brian et al. "The Role of 3D Exoscope Systems in Neurosurgery: An Optical Innovation." Cureus vol. 13,6 e15878. 23 Jun. 2021, doi:10.7759/cureus.15878

[21] Langer, David J et al. "Advances in Intraoperative Optics: A Brief Review of Current Exoscope Platforms." Operative neurosurgery (Hagerstown, Md.) vol. 19,1 (2020): 84-93. doi:10.1093/ons/opz276

[22] Oertel, Joachim M, and Benedikt W Burkhardt. "Vitom-3D for Exoscopic Neurosurgery: Initial Experience in Cranial and Spinal Procedures." World neurosurgery vol. 105 (2017): 153-162. doi:10.1016/j.wneu.2017.05.109

[23] Muhammad, Sajjad et al. "Preliminary experience with a digital robotic exoscope in cranial and spinal surgery: a review of the Synaptive Modus V system." Acta neurochirurgica vol. 161,10 (2019): 2175-2180. doi:10.1007/s00701-019-03953-x

[24] Jason Geng, "Structured-light 3D surface imaging: a tutorial," Adv. Opt. Photon. 3, 128-160 (2011). Doi:10.1364/AOP.3.000128

[25] Huang, Xuanlun et al. "Polarization structured light 3D depth image sensor for scenes with reflective surfaces." Nature communications vol. 14,1 6855. 27 Oct. 2023, doi:10.1038/s41467-023-42678-5

[26] Meana, Victor et al. "Laser Triangulation Sensors Performance in Scanning Different Materials and Finishes." Sensors (Basel, Switzerland) vol. 24,8 2410. 10 Apr. 2024, doi:10.3390/s24082410

[27] M. Levoy, "Light Fields and Computational Imaging," in Computer, vol. 39, no. 8, pp. 46-55, Aug. 2006, doi: 10.1109/MC.2006.270.

[28] Zhou, S., Zhu, T., Shi, K. et al. Review of light field technologies. Vis. Comput. Ind. Biomed. Art 4, 29 (2021). doi:10.1186/s42492-021-00096-8

[29] Jeon, Hae-Gon & Park, Jaesik & Choe, Gyeongmin & Park, Jinsun & Bok, Yunsu & Tai, Yu-Wing & Kweon, Inso. (2015). Accurate Depth Map Estimation from a Lenslet Light Field Camera. Doi:10.1109/CVPR.2015.7298762.

[30] Skirboll, S S et al. "Functional cortex and subcortical white matter located within gliomas." Neurosurgery vol. 38,4 (1996): 678-84; discussion 684-5.

[31] Senders, Joeky T et al. "Agents for fluorescence-guided glioma surgery: a systematic review of preclinical and clinical results." Acta neurochirurgica vol. 159,1 (2017): 151-167. doi:10.1007/s00701-016-3028-5

[32] Bian, Y., Ourselin, S., & Xie, Y. (2025). A light field fluorescence imaging system for neurosurgery: Proc.SPIE. doi:10.1117/12.3042419

[33] Rangappa, S., Matharu, R., Petzing, J. et al. Establishing the performance of low-cost Lytro cameras for 3D coordinate geometry measurements. Machine Vision and Applications 30, 615–627 (2019). doi:10.1007/s00138-019-01013-z

[34] Bian, Y., Gil, H. M., Elliot, M., Stasiuk, G. J., Shapey, J., Vercauteren, T., Ourselin, S., & Xie, Y. (2025). Solid tissue-mimicking phantoms for PpIX fluorescence imaging with improved fluorescence photostability: Proc.SPIE. doi:10.1117/12.3040890

[35] Li, Yifeng et al. "Design of a Novel Microlens Array and Imaging System for Light Fields." Micromachines vol. 15,9 1166. 21 Sep. 2024, doi:10.3390/mi15091166

[36] Pitskhelauri, D I et al. "Craniotomy size determines the neurosurgeon - microscope interaction: A proof-of-concept study." Journal of clinical neuroscience : official journal of the Neurosurgical Society of Australasia vol. 112 (2023): 48-54. doi:10.1016/j.jocn.2023.04.008

[37] Rösler, Judith et al. "Clinical implementation of a 3D4K-exoscope (Orbeye) in microneurosurgery." Neurosurgical review vol. 45,1 (2022): 627-635. doi:10.1007/s10143-021-01577-3

[38] Christopher Hahne, Amar Aggoun. "PlenoptiSign: An optical design tool for plenoptic imaging." SoftwareX, Volume 10, 2019, 100259, ISSN 2352-7110. Doi:10.1016/j.softx.2019.100259.

[39] Mahmoudpour, S., Pagliari, C. & Schelkens, P. Learning-based light field imaging: an overview. J Image Video Proc. 2024, 12 (2024). doi:10.1186/s13640-024-00628-1